\documentclass{mem}
\usepackage{natbib}
\usepackage{txfonts}
\usepackage{balance}
\usepackage{graphicx}
\idline{75}{1}
\begin{document}
\def\teff{$T\rm_{eff }$}
\def\kms{$\mathrm {km s}^{-1}$}

\title{On the He burning phases of the Carina dSph}

   \subtitle{}

\author{
M.\,Fabrizio\inst{1} 
\and A.\,Pietrinferni\inst{1}
\and G.\,Bono\inst{2,3}
\and P.B.\,Stetson\inst{4}
\and A.R.\,Walker\inst{5}
\and R.\,Buonanno\inst{1,2}
\and S.\,Cassisi\inst{1}
\and I.\,Ferraro\inst{3}
\and G.\,Iannicola\inst{3}
\and M.\,Monelli\inst{6}
\and M.\,Nonino\inst{7}
\and L.\,Pulone\inst{3}
\and F.\,Thev\'{e}nin\inst{8}
}

\offprints{Michele Fabrizio\\ \email{fabrizio@oa-teramo.inaf.it}}

\institute{
INAF--OATe, via M. Maggini -- 64100, Teramo, Italy
\and
Univ. Rome "Tor Vergata", Via della Ricerca Scientifica, 1 -- 00133, Roma, Italy
\and
INAF--OAR, via Frascati 33 -- 00040, Monte Porzio Catone (RM), Italy
\and
DAO--HIA, NRC, 5071 West Saanich Road, Victoria, BC V9E 2E7, Canada
\and
NOAO--CTIO, Casilla 603, La Serena, Chile
\and
IAC, Calle Via Lactea, E38200 La Laguna, Tenerife, Spain
\and 
INAF--OAT, via G.B. Tiepolo 11 -- 40131, Trieste, Italy
\and 
Obs. C\^{o}te d'Azur, BP 4229 -- 06304, Nice, France
}

\authorrunning{M. Fabrizio}

\titlerunning{On the He burning phases of the Carina dSph}

\abstract{
We performed a detailed comparison between predicted He burning phases 
and multiband photometry of the Carina dwarf spheroidal galaxy. 
We found a good agreement with the predictions computed assuming 
an $\alpha$-enhanced chemical mixture, indicating a mean metallicity 
$\rm{[Fe/H]}\sim-1.8$ with a raw observed peak-to-peak spread in iron 
abundance of 0.4$\pm$0.2~dex.

}
\maketitle{}

\section{Introduction}

Dwarf galaxies and the newly discovered ultra-faint-dwarf galaxies are
the crossroad of several theoretical and empirical investigations. The
reasons are manifold. 
\textit{a)} These relatively small stellar systems fix the limit beyond
which the dark matter halos do not form stars. This means that their
number and intrinsic properties can provide firm constraints on the
timescale during which these subhalos have been accreted by their parent
galaxies \citep{rocha12,rocha13}, and in turn to validate current dark
matter simulations of galaxy formation.   
%
%
\textit{b)} Photometric and spectroscopic data indicate that
dwarf galaxies do obey to a "linear" relation between metallicity and
luminosity (see Fig.~7 in \citealt{mateo98}). This evidence was
originally brought forward by \citet{skillman89} and cover more than
3~dex in metallicity and $\approx$15~mag in luminosity. The above
empirical evidence suggests that the chemical enrichment of these
systems was mainly driven by internal mechanisms. However, we still lack
firm constraints on the intrinsic metallicity dispersion and their possible
correlation with intermediate-age and old stellar populations.    
\textit{c)} The relevant number of dwarf galaxies recently discovered by
the SDSS in the Local Group \citep{mcconnachie12} improved the
observational basis of the so-called "density--morphology" relation.
Empirical evidence indicates that dwarf galaxies located closer to the
two giant galaxies (Milky Way, M31) are typically gas--poor, spherical
and non rotating, while those located at larger distances are gas--rich,
irregular and rotating \citep{grcevich09}. Numerical simulations suggest
that the morphology segregation might be explained as the consequence of
both ram pressure stripping and tidal stirring: once field dwarf
irregulars are accreted they will be transformed into dwarf spheroidals 
\citep[dSphs,][]{lokas10}. However, the above scenario is hampered by the 
limited sample of spectroscopic data available for old tracers in nearby 
dwarf galaxies. 

We present preliminary results on the mean metallicity of
the stellar populations in the Carina dSph using multiband photometry. 

\section{Optical Color-Magnitude diagrams}

We have already collected a large set (4152) of multi-band optical ($U$,
$B$, $V$, $I$) images of the Carina dSph \citep{bono10,stetson11}. The
optical ($V$, $B-I$; $V$, $B-V$) Carina Color-Magnitude Diagram (CMD)
has already been discussed in several investigations. On the basis of
$BV$ photometry \citet{battaglia12} found evidence of a spread in age,
when moving from the innermost regions to regions located beyond the
nominal truncation radius. 

The advanced evolutionary properties of low- and intermediate-mass stars
provide at least three independent metallicity indicators: \textit{a)}
the slope and the spread in color of the red giant branch (RGB);
\textit{b)} the spread in color of red clump (RC) stars; \textit{c)} the
zero-age-horizontal-branch (ZAHB) luminosity and the spread in magnitude
of horizontal branch (HB) stars. The photometric analysis showed a very limited
range in color covered  by RGB and RC stars, as well as the spread in
magnitude of blue and red HB stars\footnote{The off-ZAHB evolution 
also increases the spread in magnitude of HB stars. However, a detailed 
analysis of the pulsation properties of Carina RR Lyrae indicates that 
they are marginally affected by evolutionary effects \citep{coppola13}.}.
The above evidence suggests that Carina stars cover a range in metallicity 
smaller than $\sim0.6$~dex \citep{bono10}.

A similar analysis was also performed by \citet{stetson11},  but using
stellar isochrones of different ages and chemical compositions. The
theoretical framework they adopted for both scaled-solar and
$\alpha$-enhanced evolutionary models is based on isochrones available
in the BaSTI\footnote{\tt http://albione.oa-teramo.inaf.it/} database
\citep{pietrinferni04,pietrinferni06}. They found that both the old (12~Gyr)
and the intermediate-age (4-6~Gyr) sub-populations show a limited
spread in chemical composition. The above results support mean iron
abundances ($\rm{[Fe/H]}=-1.72$, $\sigma=0.24$) based on high-resolution
spectra for 44 RGs recently provided by \citet[][see also
\citealt{lemasle12} and \citealt{venn12}]{fabrizio12}. Stellar
populations in Carina were also investigated using synthetic CMDs and a
new Bayesan approach by \citet{small13}. They found that the 34\%\ of
the stars are old (13 Gyr), while the 46\%\ are intermediate (4.5-7
Gyr), plus a plume of younger objects.    

We discuss the evolutionary properties of the Carina
sub-populations using HB stars are tracers of the old population
and RC  stars as tracers of the intermediate-age population.   

\begin{figure*}[]
\begin{flushright}
\includegraphics[width=0.93\textwidth]{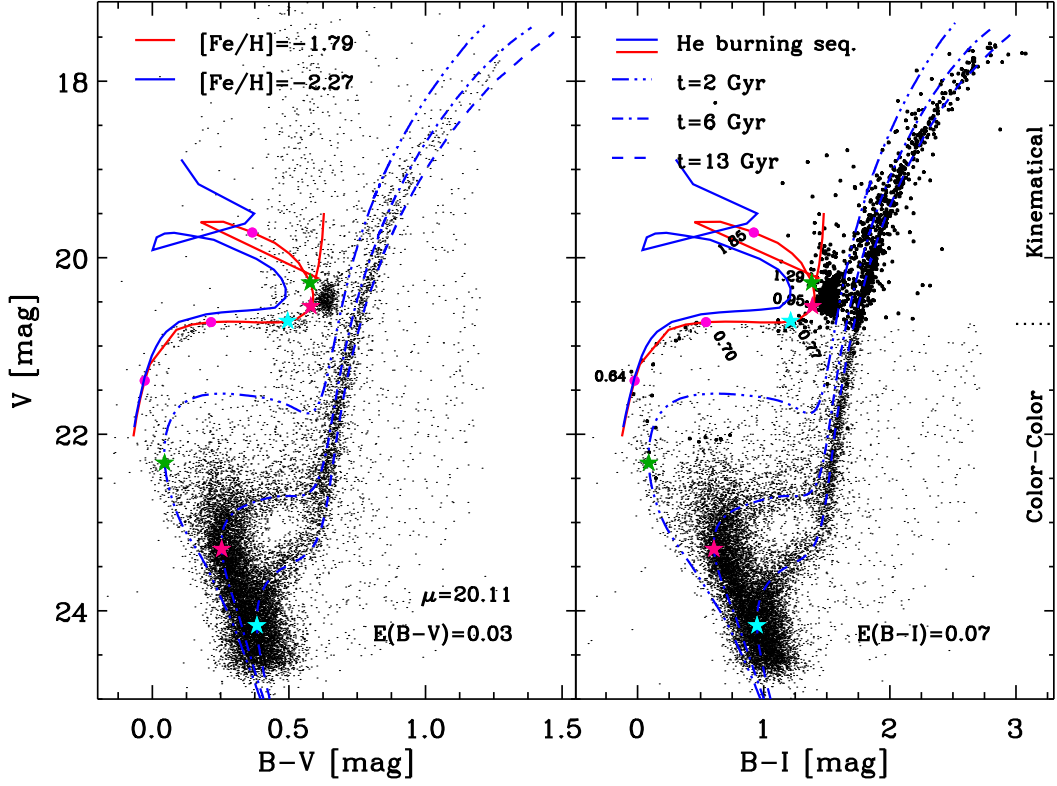}
\end{flushright}
\vspace*{-8.6truecm}
\caption{\footnotesize Left -- Comparison in the $V$, $B-V$ CMD between
candidate Carina stars selected on the basis of the $U-V$, $B-I$
color-color plane \citep{bono10} and isochrones computed adopting a
scaled-solar chemical composition ($\rm{[Fe/H]}=-1.79$). The ages range
from 2 (dashed-dotted-dotted) to 13 (dashed)~Gyr. The solid blue and
red lines show the predicted He burning sequences for two different iron
abundances. The adopted true distance modulus and reddening are also
labelled. 
Right -- same as the left, but for the $V$, $B-I$ CMD. The 1379 stars
with visual magnitudes brighter than 20.75 were selected according to
kinematic properties \citep{fabrizio11}. The labels along the
metal-poor He burning sequence display selected mass values (see text
for details).}
\label{solar}
\end{figure*}
%
The Fig.~\ref{solar} shows the comparison between observations and
theory, in the $V$, $B-V$ (left) and $V$, $B-I$ (right) CMD. To compare
theory and observations we adopted the distance modulus and the
reddening available in the literature
\citep{monelli03,dallora03,pietrzynski09}. The selective extinction
coefficients that we adopted are the following: $R_V=3.07$, $R_B=4.07$
and $R_I=1.65$ from \citet[][see also \citealt{bono10b} for further
details]{mccall04}. Candidate Carina stars were selected according to
the $U-V$, $B-I$ color-color plane \citep{bono10}. In the right panel,
the 1379 stars with visual magnitudes brighter than 20.75 were selected
according to kinematic properties \citep{fabrizio11}. The isochrones
range from 2 (dashed-dotted-dotted) to 13 (dashed)~Gyr. They have been
computed assuming an iron abundance of $\rm{[Fe/H]}=-1.79$ and a
scaled-solar chemical mixture. 
The precision of the photometric catalog allows us to constrain the
properties of the different sub-populations for the old, the
intermediate-age and the young stellar populations. The old population
is well reproduced by an isochrone of 13~Gyr, this means a main sequence
turn off (MSTO) mass value of $M=0.77$~M$_\odot$. An isochrone of 6~Gyr
is a good average for the intermediate-age population, with a MSTO mass
of $M=0.95$~M$_\odot$. Note that this star formation event shows
evidence of a spread in age of the order of 2~Gyr. The blue plume
(youngest stellar population) is well reproduced by a 2~Gyr isochrone
(MSTO mass $M=1.29$~M$_\odot$), once again with an age spread. The BaSTI
evolutionary tracks were computed assuming a Reimers mass loss rate
($\eta=0.4$), this means that the stellar structures approaching the tip
of the RGB (TRGB) will have smaller mass values: 0.69 (old), 0.89
(intermediate) and 1.25 (young)~M$_\odot$. The relative difference is
due to the fact that the RGB phase of the old population is
significantly longer (440 Myr) when compared with the intermediate (250
Myr) and the young (120 Myr) stellar population. Indeed, low-mass stars 
develop an electron degenerate He core, and therefore, the He ignition 
(3$\alpha$) is delayed until the He core mass approaches 0.5~M$_\odot$.

The solid blue and red lines display predicted He burning sequences of
old and intermediate-age progenitors for two different metal abundances
(see labeled values). The ZAHB of the old population was 
computed\footnote{The ZAHB $V$-magnitude was increased by 0.05~mag
to account for the decrease in the He-core mass caused by 
the use of more updated conductive opacities \citep{cassisi07}.} assuming an age for the
progenitor of 13~Gyr, which means a mass value at the TRGB of
0.69~M$_\odot$. The position along the ZAHB of a structure with 
the mass value of the MSTO was marked with a cyan star.
The above evolutionary scenario only accounts for "reddish" HB stars.
However, empirical evidence indicates that the HB morphology ranges from
extremely hot ($\approx$35,000~K), less massive (0.55~M$_\odot$) to cool
($\approx$5,000~K), more massive (0.80~M$_\odot$) HB stars. To account
for this difference in HB morphology we assume that hot HB stars
experienced a more efficient mass loss along the RGB. This assumption is
supported by empirical evidence, suggesting that the mass-loss rate is
far from being steady along the RGB \citep{mcdonald11}. To provide a 
more quantitative analysis of this phenomenon we also marked the 
ZAHB stellar structures with a mass of 0.64 and 0.70~M$_\odot$, 
i.e. structures that lost along the RGB the
$\sim$17\%\ and the $\sim$9\%\ of their initial mass. 

The intermediate-age He burning sequence (IAHBS) becomes
brighter and slightly hotter. The change is mainly caused by the fact
that in stellar structures more massive than $\sim$1.3~M$_\odot$ the
He burning takes place in a He core that is only partially affected
by electron degeneracy (RC). 
For these stellar structures, the decrease in He burning efficiency, 
due to the decrease in the He core mass, is superseded  by the 
increased efficiency of the hydrogen burning shell.
The magenta star marks along the IAHBS the position of the
6~Gyr stellar population. Stellar structures with slightly younger age
show a trend similar to RC stars: they become steadily brighter and
hotter (green star). A sharp change in this behavior takes place for 
$M\ge2.0$~M$_\odot$ ($t\sim600$~Myr), since they become suddenly
fainter and cooler. Structures more massive than this limit undergo 
a quiescent He burning ignition at the tip of the RGB 
\citep{sweigart89,fiorentino12}. 
The He burning phase of even more massive structures develop, for the 
above abundance, the so-called blue loops \citep{bono00}. 
This scenario is supported by the identification in
Carina of several Anomalous Cepheids associated to the
RC phase and a few candidate short-period classical Cepheids,
associated to the blue loop \citep{coppola13}. 

The comparison between theory and observations shows that both  
metal-poor and metal-rich predictions are either
brighter (ZAHB) or hotter (RC) than the observed HB and RC stars. 
%
\begin{figure*}[]
\vspace*{-9.5truecm}
\begin{flushright}
\includegraphics[width=0.93\textwidth]{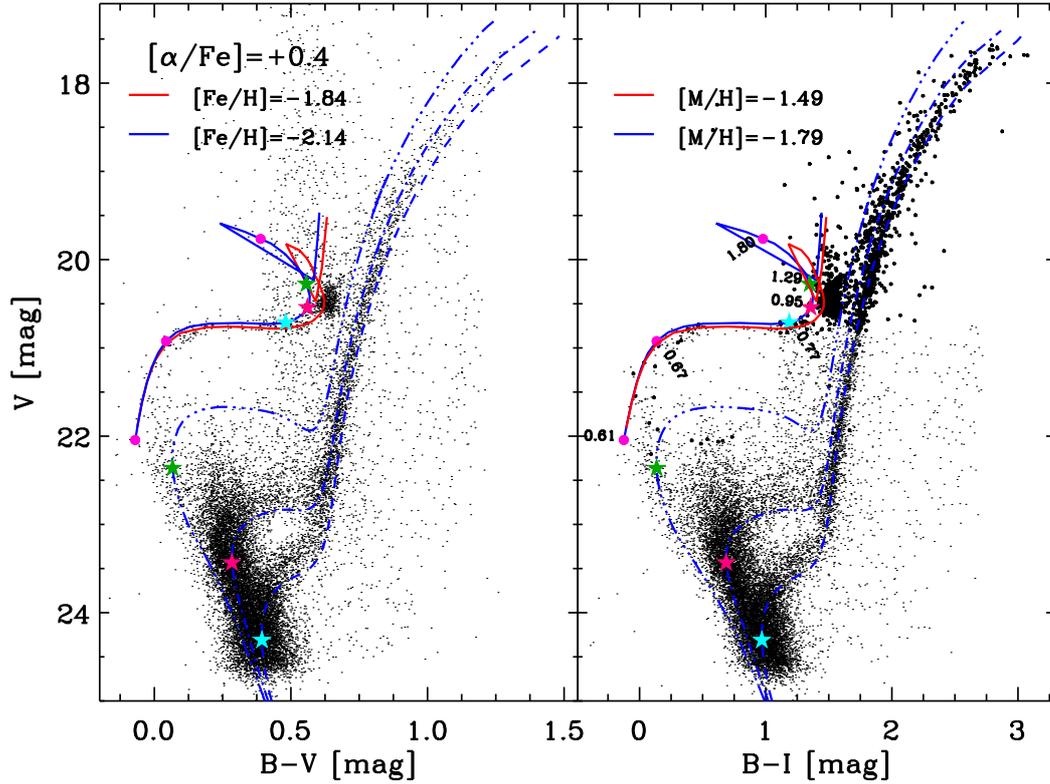}
\end{flushright}
\vspace*{0.7truecm}
\caption{\footnotesize Same as Fig.~\ref{solar}, but for isochrones and
He burning sequences constructed assuming an $\alpha$-enhanced chemical
mixture. The iron abundances and the global metallicities [M/H] are also
labelled.}
\label{alpha}
\end{figure*}
%
We performed a similar comparison using
evolutionary prescriptions constructed by assuming an  $\alpha$-enhanced
chemical mixture $\rm{[\alpha/Fe]}=+0.4$. Data plotted in
Fig.~\ref{alpha} show that $\alpha$-enhanced isochrones and He  
burning sequences agree quite well with observations. The comparison also
indicates that old and intermediate-age He burning stars
could be explained with a raw observed peak-to-peak spread in iron 
abundance of 0.4$\pm$0.2~dex. 
However, the evidence that $\alpha$-enhanced predictions
agree better with observations than scaled-solar ones is at odds with
current spectroscopic measurements suggesting that Carina RGs do not 
show evidence of $\alpha$-enhancement
($\rm{[Ca+Mg+Ti/3Fe]}=0.06$, \citealt{shetrone03};
$\rm{[Ca+Mg+Ti/3Fe]}=0.07$, \citealt{venn12}).

\section{Summary}
We compared theoretical prescriptions for He burning phases with 
Carina accurate multiband photometry \citep{bono10}. The
predicted He burning sequences, from BaSTI data base, span from old to
intermediate-age progenitors. In particular, we presented the comparison
with two metal abundances and two different chemical mixtures
(scaled-solar, $\alpha$-enhanced). We found a satisfactory agreement
with the $\alpha$-enhanced chemical composition
$\rm{[\alpha/Fe]}=+0.4$, providing a mean metallicity value of
$\rm{[Fe/H]}\sim-1.8$ with a raw observed peak-to-peak spread in iron 
abundance of 0.4$\pm$0.2~dex.

\begin{acknowledgements}
This work was partially supported by PRIN-INAF 2011 "Tracing the
formation and evolution of the Galactic halo with VST" (PI: M. Marconi)
and by PRINÐMIUR (2010LY5N2T) "Chemical and dynamical evolution of the
Milky Way and Local Group galaxies" (PI: F. Matteucci).
\end{acknowledgements}

\bibliographystyle{aa}

\end{document}